\documentclass[runningheads]{llncs}
\usepackage[T1]{fontenc}

\usepackage{amsmath}
\usepackage{color}
\usepackage[dvipsnames]{xcolor}
\usepackage{xspace}
\usepackage{hyperref}
\usepackage{url}
\usepackage{subcaption}
\usepackage{listings}
\usepackage{ifthen}

\usepackage{pgf,tikz}
\usepackage{amssymb}
\usepackage{amstext}
\usepackage{floatrow}

\usepackage[group-separator=\text{\,},
            group-minimum-digits=4,
            detect-all]{siunitx}
\usepackage[smaller,nolist,nohyperlinks]{acronym}

\usepackage{csquotes} 
\usepackage{chngcntr}

\renewcommand{\epsilon}{\varepsilon}
\renewcommand{\phi}{\varphi}

\DeclareGraphicsExtensions{.pdf,.jpeg,.png}
\graphicspath{{images/}}

\newboolean{anonymous} 
\setboolean{anonymous}{false}
\newcommand{\onlyan}[1]{\ifthenelse{\boolean{anonymous}}{#1}{}}
\newcommand{\onlynonan}[1]{\ifthenelse{\boolean{anonymous}}{}{#1}}

\newboolean{long} 

\setboolean{long}{false}

\newcommand{\lvsv}[2]{\ifthenelse{\boolean{long}}{#1}{#2}}

\newcommand{\textsff}[1]{\textsf{#1}}

\newcommand{\Eg}{E.\,g.\@\xspace}
\newcommand{\eg}{e.\,g.\@\xspace}

\newcommand{\ie}{i.\,e.\@\xspace}
\newcommand{\etal}{et al.\@\xspace}

\lstdefinelanguage{EBNF}{
  numbers=none,
  stringstyle=\color{OliveGreen},
  morestring=[b]',
  morestring=[b]"
}

\lstdefinelanguage{SCCharts}
{language=C,
	morekeywords={Pr,
		_,
		abort,
		auto,
		bool,
		call,
		clock,
		combine,
		conflict,
		confluent,
		connector,
		const,
		dataflow,
		deferred,
		delayed,
		do,
		during,
		entry,
		exit,
		expression,
		extern,
		final,
		float,
		for,
		global,
		go,
		history,
		host,
		if,
		immediate,
		initial,
		input,
		int,
		is,
		join,
		label,
		max,
		min,
		nondeterministic,
		none,
		once,
		output,
		period,
		pre,
		preceding,
		print,
		pure,
		random,
		randomize,
		ref,
		region,
		reset,
		scchart,
		schedule,
		scope,
		shallow,
		signal,
		state,
		static,
		string,
		succeeding,
		suspend,
		to,
		undefined,
		unsigned,
		val,
		violation,
		weak,},
	moredelim=**[is][\color{red}]{@}{@},
	breaklines=true,  
	tabsize=2,
	breakindent=2mm,
	breakatwhitespace=true,
	columns=fullflexible,
	backgroundcolor=\color{white!10}
}

\lstdefinelanguage{LF}
{language=C,
	morekeywords={Pr,
		target,
		logical,
		physical,
		action,
		reaction,
		startup,
		main,
		reactor,
		timer,
		state,
		new},
	moredelim=**[is][\color{red}]{@}{@},
	breaklines=true,  
	tabsize=2,
	breakindent=2mm,
	breakatwhitespace=true,
	columns=fullflexible,
	backgroundcolor=\color{white!10}
}

\usepackage{graphicx}
\usepackage{color}

\setboolean{anonymous}{false}
\begin{acronym}[KIELER]
    \acro{lf}[LF]{Lingua Franca}
    \acro{ono}[ONO]{Obviously Non-Optimal}
    \acro{sb}[S\&B]{Scheidt \& Bachmann System Technik GmbH}
\end{acronym}
    
\acused{api}
\acused{css}
\acused{kieler}
\begin{document}
\counterwithout{lstlisting}{chapter}
\title{Diagram Control and Model Order for Sugiyama Layouts}
%
%
\onlynonan{
    \author{S\"oren Domr\"os\orcidID{0000-0002-8011-8484} \and 
    Reinhard~von~Hanxleden\orcidID{0000-0001-5691-1215}}
    \authorrunning{S. Domr\"os and R. von Hanxleden}
    %
    \institute{Department of Computer Science, Kiel University, Kiel, Germany\\
        \email{\{sdo,rvh\}@informatik.uni-kiel.de}}
}
\onlyan{
    \author{First Author\inst{1}\orcidID{0000-0000-0000-0000} \and
		Second Author\inst{1}\orcidID{0000-0000-0000-0000} \and
		Third Author\inst{1}\orcidID{0000-0000-0000-0000}
	}
	\authorrunning{Author et al.}
	%
	\institute{
		Institution \email{mail@mail.mail}
	}
}
\maketitle              
\begin{abstract}

Graphical \textsc{wysiwyg} editors programming languages are popular since they allow to \emph{control} the diagram layout to express \emph{intention} via \emph{secondary notation} such as proximity and topology.
However, such editors typically require users to do manual layout.
Conversely, automatic layout of diagrams typically fails to capture intention because graphs are usually considered to not contain any order.
\emph{Model order} can combine the desire for control of secondary notation with automatic layout, without additional overhead, since the textual model already employs secondary notation.
We illustrate how model order can exert control on the example of programming languages SCCharts and Lingua Franca.
We also propose a first guidebook how such model order configurations can be extracted for other programming languages with a graphical notation.

\end{abstract}
\section{Introduction}

Automatic layout rises in popularity, as seen in the example of elkjs\footnote{\url{https://npmtrends.com/elkjs}} with, as of this writing, more than 600.000 weekly downloads.
Even though automatic layout improved over time and gets more widely used, \textsc{wysiwyg} editors that rarely employ automatic layout are still very common.
\textsc{wysiwyg} editors place the burden of layout on the user, which can be a severe impediment to productivity~\cite{Petre95}.
However, \textsc{wysiwyg} editors have the advantage that they allow controlling secondary notation~\cite{Petre95} by creating order, grouping, or alignment in a very direct way and on a graphical level, which is desirable.
As Taylor reported for \textsc{wysiwyg} type-setting~\cite{Taylor96}: \enquote{People like having feedback and control.}

One approach to augment automatic layout of diagrams with control is to let the user formulate explicit layout constraints, \eg, through (textual) model annotations or via some \textsc{wysiwyg}-like graphical interaction~\cite{DwyerMW09b,PetzoldDSvH23}.
Constraints have the advantage to be integrated into layout algorithms.
Hence, layout does not need to be done on a pixel granularity, which when done manually often has inconsistencies \cite{PurchaseAKN+20}.
Instead, they focus on topology, alignment, or proximity of nodes.
Constraints, however, require additional effort beyond creating a textual model, and sometimes require knowledge about the underlying layout algorithm to be used effectively, as reported by users and developers of \acs{elk}~\cite{DomroesvHS+23}.

In this paper, we investigate the research question how to exert control by using \emph{model order}~\cite{DomroesRvH23} for \emph{Sugiyama or layered layouts}~\cite{SugiyamaTT81} whenever the textual model in a tool that employs modeling pragmatics~\cite{vonHanxledenLF+22}, by using a textual model and a graphical model side-by-side, expresses secondary notation (R1).
We also explore how model order integrates with the mental map, layout stability, aesthetic criteria, and secondary notation in text and diagram (R2).
Since model order integrates fully into automatic layout, we can use the best of both a textual and a graphical model \cite{vonHanxledenLF+22}, without the overhead that is
introduced by layout constraints or manual layout in \textsc{wysiwyg} editors~\cite{Petre95}.

\begin{figure}[tb]
	\centering
    \begin{subfigure}[b]{0.27\textwidth}
        \centering
        \begin{lstlisting}[language=SCCharts, escapeinside={(*}{*)}, abovecaptionskip=0pt, captionpos=t]
scchart Example1a {
    (*$\dots$*)
    initial state Start {
        (*$\dots$*)
    }
    if (*$\dots$*) go to Send
    if (*$\dots$*) go to Receive
    state Send { (*$\dots$*) }
    (*$\dots$*)
    state Receive { (*$\dots$*) }
    (*$\dots$*)
    state Done { (*$\dots$*) }
    immediate go to Start
}
        \end{lstlisting}
        \caption{}
        \label{fig:send-receive-text}
    \end{subfigure}
    \begin{subfigure}[b]{0.21\textwidth}
        \centering
        \includegraphics[scale=0.4]{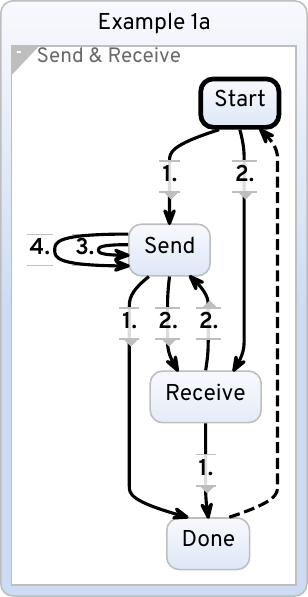}
        \caption{}
        \label{fig:send-receive_goodorder}
    \end{subfigure}
    \begin{subfigure}[b]{0.27\textwidth}
        \centering
        \begin{lstlisting}[language=SCCharts, escapeinside={(*}{*)}, abovecaptionskip=0pt, captionpos=t]
scchart Example1b {
    (*$\dots$*)
    initial state Start {
        (*$\dots$*)
    }
    if (*$\dots$*) go to Send
    if (*$\dots$*) go to Receive
    state Receive { (*$\dots$*) }
    (*$\dots$*)
    state Send { (*$\dots$*) }
    (*$\dots$*)
    state Done { (*$\dots$*) }
    immediate go to Start
}
        \end{lstlisting}
        \caption{}
        \label{fig:send-receive-text2}
    \end{subfigure}
    \begin{subfigure}[b]{0.21\textwidth}
        \centering
        \includegraphics[scale=0.4]{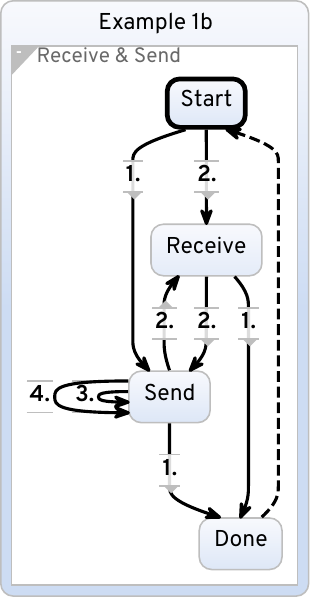}
        \caption{}
        \label{fig:send-receive_badorder}
    \end{subfigure}
	\caption{
        Two semantically identical variants of an obfuscated SCCharts model created by and used with permission of \acl{sb}.
        The textual source models can be seen in \autoref{fig:send-receive-text} and \ref{fig:send-receive-text2} with the matching diagram in \autoref{fig:send-receive_goodorder} and \autoref{fig:send-receive-text2}.
    }
	\label{fig:send-receive}
\end{figure}

The Sugiyama algorithm consisting of the phases cycle breaking, layer assignment, crossing minimization, node placement, and edge routing may produce drawings as the one depicted in \autoref{fig:send-receive}.
Here, nodes are assigned to (horizontal) layers such that nodes in the same layer are not connected by an edge.
The cycle breaking step 
determines which edges should go against the layout direction (here the layout direction is top-to-bottom) and hence determines the vertical order of \textsff{Send} and \textsff{Receive}.
Blindly optimizing aesthetic criteria disregards the intention of the model and leaves no leverage to control the layout.
The two versions are semantically identical, the vertical ordering in \autoref{fig:send-receive_goodorder} suggests that \textsff{Send} happens before \textsff{Receive} while \autoref{fig:send-receive_badorder} suggests the inverse.
\autoref{fig:send-receive_goodorder} was created using the textual model in \autoref{fig:send-receive-text}, while \autoref{fig:send-receive_badorder} was created using \autoref{fig:send-receive-text2}.
Even though it would be possible to set layout constraints for the textual model in \autoref{fig:send-receive-text2} to create the drawing in \autoref{fig:send-receive_goodorder}, we suggest changing the textual order instead, as we argued here,
since the mental map, the inner representation of the model, and secondary notation of the textual model should not diverge from the mental map and the secondary notation of the diagram.

To answer the research question R1 and R2 stated above, we contribute
\begin{itemize}
    \item an investigation how controlling the diagram via model order influences the mental map, layout stability, and secondary notation in \autoref{sec:concept},
    \item an analysis of SCCharts and \ac{lf} in the context of model order and control in \autoref{sec:configuration}, and
    \item a guidebook how to extract order information in \autoref{sec:feedback-gathering}.
\end{itemize}
\autoref{sec:conclusion} summarizes and generalizes ours insights and suggests future research.


\section{Related Work}
\label{sec:related-work}

Purchase \etal \cite{PurchasePP12,Purchase14} investigated what aesthetic criteria humans adhere to when drawing graphs given by a textual description.
As also stated by Purchase \etal, other studies \cite{Purchase97,Sugiyama02,HuangHE07,HuangEHL13} focus on reduction of edge crossings, symmetry, placement of important nodes at the top, large angles between incident edges, and average edge length.
Purchase \etal investigated whether there are additional criteria people favor when drawing graphs, by analyzing the final drawings and the intermediate steps of two graphs created by the participants.
The study revealed that people prefer to place nodes on a grid and that they initially use aesthetic criteria such as only vertical or horizontal edges or nodes ordered lexicographically  or by their occurrence in the graph representation.

A follow-up~\cite{Purchase14} study further investigated why the participants revised their drawing to no longer conform to initially used aesthetic criteria, whether they prefer a grid-based layout algorithm, and what influence the graph representation had in favoring a grid-based layout.
The study revealed that the participants just worked through the graph node by node or edge by edge depending on the graph representation.
In the first study, they also used the lexicographic order, since the labeling of the nodes A to J suggests that this order is intentional.
The follow-up study used $\{P \dots Z \}\backslash \{O,Q\}$ as node names for their second graph and a textual graph representation that was obviously not ordered lexicographically.
Here participants used the ordering given by the representation for their initial drawing.
They intuitively used the given model order of the graph representation and revised this order partially if the result created undesired crossings or clutter.

Kieffer \etal \cite{KiefferDMW16} investigated how humans create orthogonal layouts and designed a layout algorithm that used their decisions as inspiration resulting in HOLA, an algorithm that aims to create orthogonal layouts inspired by human layouts.
Their graphs did not have any node names and did not have a textual representation but an initial drawing describing it, which might influence the mental map of the graph.
However, a textual graph description would probably also affect the human-made layouts, as it was experienced by Purchase.

Most graphs used in studies about graph drawing aesthetics are often only graphs without any intention in them and not models of a real-world application.
It only makes sense to use model order or layout constraints if the underlying model has intention that users want to express using secondary notation.
Participants of the study by Purchase \etal~\cite{PurchasePP12} deemed the order of edges, nodes, or the lexicographic order as intention.
To evaluate model order, we hence need real models and preferably the developers that built them to investigate what their intention in specific placements was.
Hence, we also need to focus on human-made models and---if applicable to the language---on realistic models, \ie, models of a real-world application.
Student models, however, are often unrealistic and their developers are often beginners who tend to create misleading secondary notation in their layouts \cite{Petre95}.
Moreover, evaluation of model order configurations for layout can only be done in an interactive context while considering the different creation steps, as Purchase \etal showed.

\section{Control and Model Order}
\label{sec:concept}

This section explores how model order relates to mental map, stability, secondary notation, control, and aesthetic criteria on a meta level (R2).
For this work, we focus on tools for model-driven-engineering that use text and diagram side-by-side utilizing the concept of modeling pragmatics~\cite{vonHanxledenLF+22}.
In this setup, controlling the layout using model order matches the mental map of the textual model with the diagram and allows us to transfer textual secondary notation into the diagram.

Typically, the \emph{mental map} describes the mental image a user has of a graphical model~\cite{PurchaseHG06}.
Misue \etal~\cite{MisueELS95} define preserving the mental map as preserving orthogonal ordering, which we call topology, \eg, nodes that are placed above of other nodes remain above of each other, and proximity, \eg, nodes that are next to each other remain next to each other.
Preserving the mental map is especially important when working with models for real use-cases since they are typically large, hierarchical, and do not fit on a single screen.
These are also the reasons why we abbreviated the models shown in this paper by filtering inner behavior.

The concept of a mental map is usually associated with diagrams or other \enquote{map-like} two-dimensional representations.
However, as we argue here, this concept  also exists in textual models.
Developers already create a mental map while creating the textual model and not only by looking at the accompanying diagram.
Hence, we might compromise the mental map if the textual model does not match the graphical model, as it would be the case if \autoref{fig:send-receive-text} would create the diagram in \autoref{fig:send-receive_badorder}.
Since the textual ordering only has one dimension but the drawing has two to express order, we have to further determine which dimension in the diagram corresponds which textual ordering for a given language.
Finding this model order configuration for a given language allows us to control the diagram using model order to visualize secondary notation.

Similarly, \emph{secondary notation} exists in textual models.
\Eg, developers begin to write the textual model with an initial state at the top, final states are typically at the bottom, and nodes that should be next to each other in the drawing are typically also placed next to each other in the textual model, as seen in \autoref{fig:send-receive}.
Since intentional secondary notation exists in the textual source, we should control the layout using the textual model order to bring secondary notation from the text into the diagram.
This requires us to analyze realistic models developed by experts that correctly employ secondary notation~\cite{Petre95} to learn how to control the two-dimensional diagram via the one-dimensional textual model.

\emph{Control} is a very desirable aspect of layout, as also reported by Taylor for \textsc{wysiwyg} type-setting~\cite{Taylor96}.
This might be the reason \textsc{wysiwyg} editors
are still implemented even though moving boxes around is a tedious process \cite{Petre95}.
Moving boxes to desired positions directly creates secondary notation.
This level of control can also be achieved using interactive constraint frameworks~\cite{DwyerMW09b,PetzoldDSvH23} with the advantage of automatic layout.
However, using model order inside a layout algorithm can similarly exert control by treating the textual ordering as a constraint.
Moreover, model order can also be used as a tie-breaker together with common aesthetic criteria creating different levels of control~\cite{DomroesRvH23}.
However, since the textual order is one-dimensional and the diagram has two dimensions, model order strategies have to be configured to avoid misleading secondary notation, which we explore for SCCharts and \ac{lf} in \autoref{sec:configuration}.

\begin{figure}[tb]
    \begin{subfigure}[b]{0.25\textwidth}
        \centering
        \begin{lstlisting}[language=LF, escapeinside={(*}{*)}, abovecaptionskip=0pt, captionpos=t]
(*$\dots$*)
l1(*$\dots$*)=(*$\dots$*)dram_storage(*$\dots$*)
l1(*$\dots$*)=(*$\dots$*)cdn_cache(*$\dots$*)
l1(*$\dots$*)=(*$\dots$*)load_balancer(*$\dots$*)

l2(*$\dots$*)=(*$\dots$*)dram_storage(*$\dots$*)
l2(*$\dots$*)=(*$\dots$*)cdn_cache(*$\dots$*)
l2(*$\dots$*)=(*$\dots$*)load_balancer(*$\dots$*)
(*$\dots$*)
        \end{lstlisting}
        \caption{}
        \label{fig:cdn-cache-snippet-text}
    \end{subfigure}

    \begin{subfigure}[b]{0.485\textwidth}
        \centering
        \includegraphics[scale=0.5]{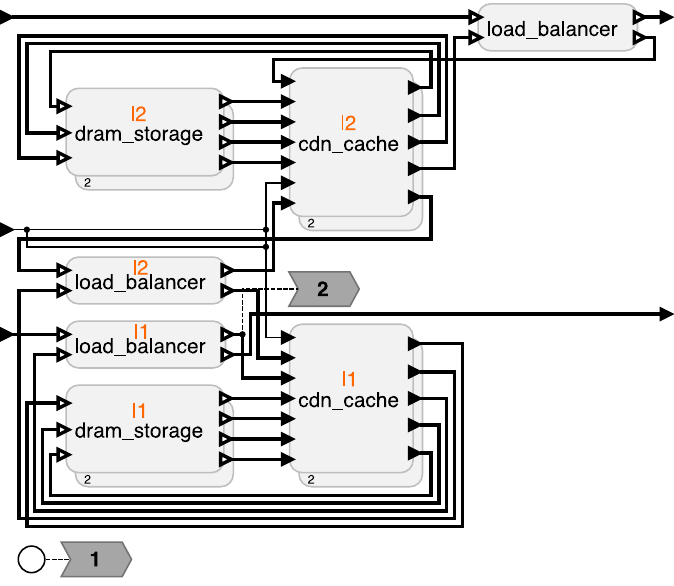}
        \caption{}
        \label{fig:cdn-cache-snippet_noorder}
    \end{subfigure}
    \begin{subfigure}[b]{0.485\textwidth}
        \centering
        \includegraphics[scale=0.5]{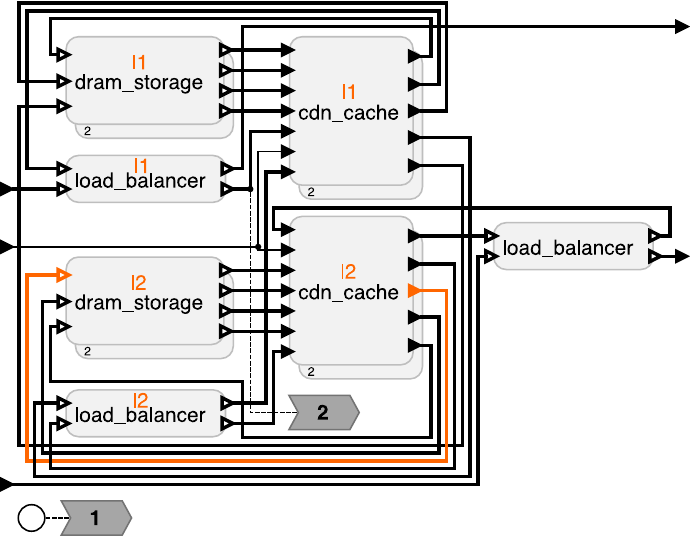}
        \caption{}
        \label{fig:cdn-cache-snippet_order}
    \end{subfigure}
    \caption{
        \autoref{fig:cdn-cache-snippet-text} shows a snippet of the inner behavior of the \textsff{Pop} reactor of the \textsff{cdn\_cache\_demo} model, with l1 above l2.
        \autoref{fig:cdn-cache-snippet_noorder} shows a layout that optimizes edge crossing, but violates the l1/l2-ordering of \autoref{fig:cdn-cache-snippet-text}.
        \autoref{fig:cdn-cache-snippet_order} shows how an alternative layout preserves this ordering.
    }
    \label{fig:cdn-cache-snippet}
\end{figure}

Additional to control, model order provides \emph{stability}.
Stability means that small changes in the textual model only result in small changes in the diagram, which helps to maintain the mental map.
Hence, introducing a new edge from \textsff{Receive} to \textsff{Send} in \autoref{fig:send-receive_goodorder} should not change the order of \textsff{Send} and \textsff{Receive} in the diagram, even though it would create an additional backward edge.
Depending on the configuration, model order can control the layout to achieve stability whenever a certain ordering expresses intentional secondary notation.
This works for the cycle breaking step as explained above and for the crossing minimization step, which determines the order of nodes in a here vertical layer, and the order of ports---the anchor points of edges---on a node, as seen in \autoref{fig:cdn-cache-snippet}.
\autoref{fig:cdn-cache-snippet_noorder} does not represent the textual ordering in \autoref{fig:cdn-cache-snippet-text} in any way.
The node order in \autoref{fig:cdn-cache-snippet_order} corresponds with the node model order.
Here, expressing secondary notation via model order results in additional edge crossings.
However, if one optimizes the layout only for edge crossings, the drawing might drastically change if a new edge is introduced or elements are expanded to show their inner behavior, which results in a different port order and typically in a completely different diagram\footnote{This can be explored interactively in the \ac{lf} playground \url{https://github.com/lf-lang/playground-lingua-franca} by opening the \textsff{cdn\_cache\_demo} model created by Magnition taken from \url{https://github.com/MagnitionIO/LF_Collaboration}.}.
Configuring the layout algorithm to use node, edge, or port model order as a constraint results in a much stabler layout but may create more edge crossings.
Previous work~\cite{DomroesvH22,DomroesRvH23} covers the effect of model order on edge crossings and backward edges.

Note that the visual clutter of edge crossings that makes it harder to follow edges, as seen in \autoref{fig:cdn-cache-snippet_order}.
Highlighting selected edges partly solves this problem by making edges traceable, as seen in \autoref{fig:cdn-cache-snippet_order}.

\section{Controlling the Layout via Model Order}
\label{sec:configuration}

To understand how a language can be controlled using model order, we have to identify what part of the textual model is intended secondary notation and how this should be transferred to the two-dimensional diagram during the cycle breaking and crossing minimization steps of the Sugiyama algorithm (R1).

\subsection{Controlling SCCharts Layout via Model Order}

SCCharts \cite{vonHanxledenDM+14} is a sequentially constructive statechart dialect that models control-flow.
\autoref{fig:send-receive} depicts an SCChart drawing with the corresponding textual model.
An SCChart consists of a declaration of inputs, outputs, constants, variables and actions, which are here filtered out of the diagram.
Moreover, it has concurrent regions (the white box called \textsff{Send \& Receive} in \autoref{fig:send-receive_goodorder}) with states and transitions between them.
The textual syntax only allows to define outgoing transitions of a state directly under the state declaration, which constraints the global edge order.
\Eg, the state \textsff{Start} has two outgoing transitions defined below it.
States might have internal behavior including everything an SCChart may consist of.

\begin{figure}
	{
		\subfloat[]{
			\centering
			\includegraphics[width=0.97\textwidth]{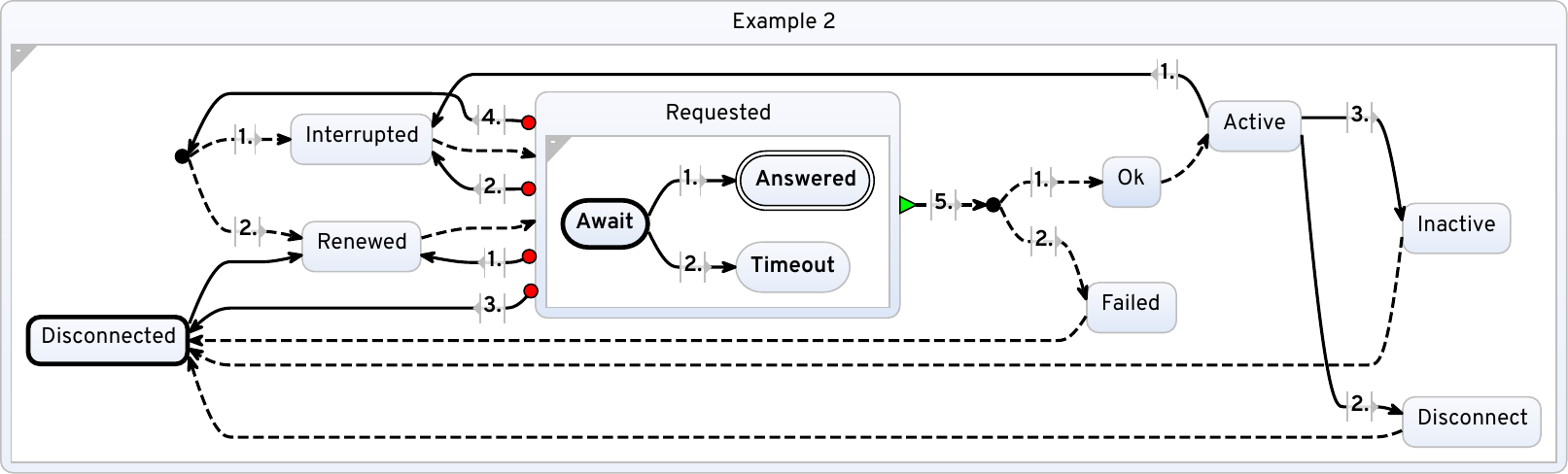}
            \label{fig:example2-connection-tie}
		}

		\subfloat[]{
			\centering
			\includegraphics[width=0.97\textwidth]{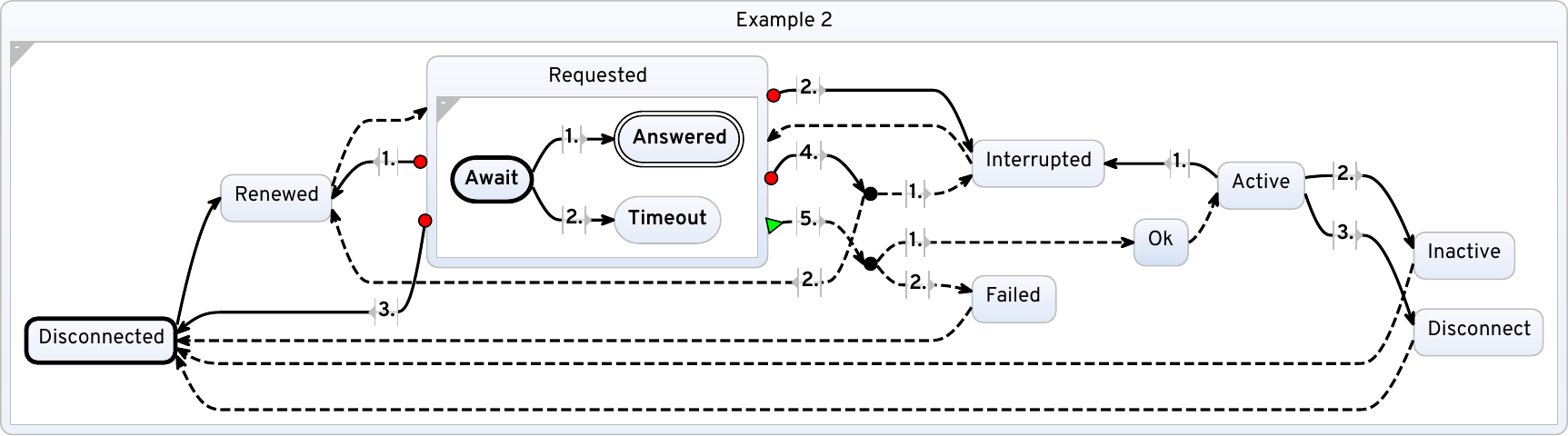}
            \label{fig:example2-connection-good}
		}

		\subfloat[]{
			\centering
			\includegraphics[width=0.97\textwidth]{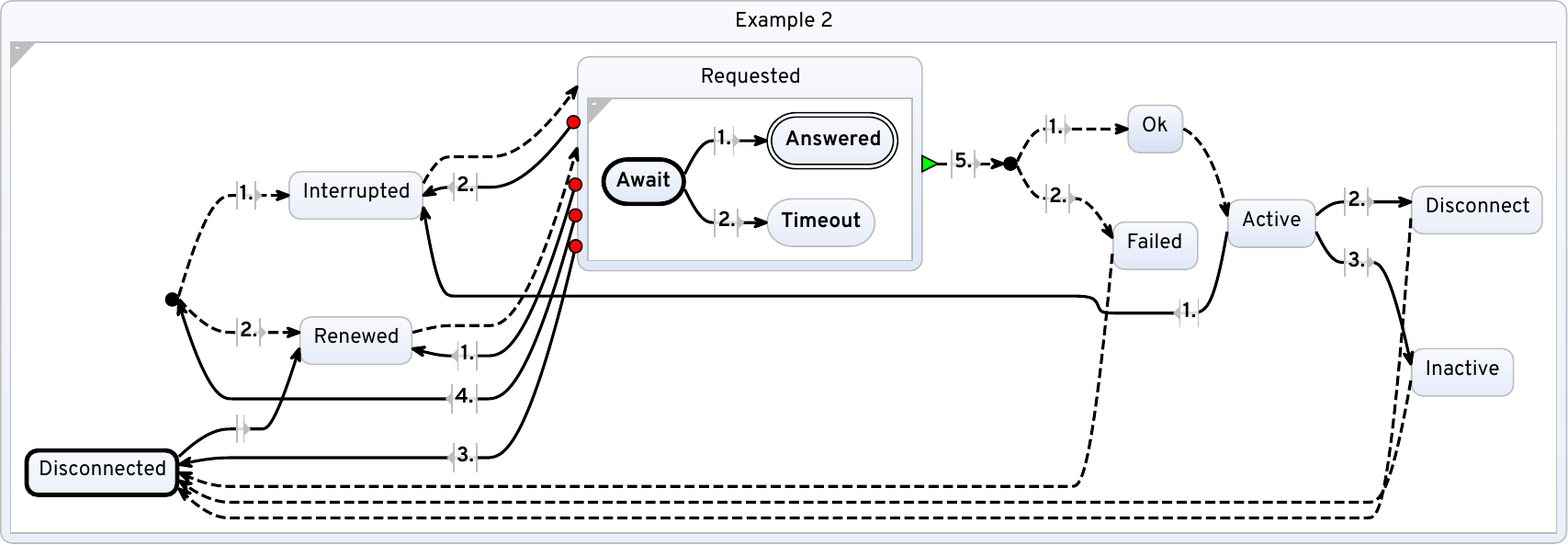}
            \label{fig:example2-connection-problem}
		}
        
	}
    \caption{An obfuscated SCCharts model created by and used with permission of \acl{sb}.
    \autoref{fig:example2-connection-tie} uses model order as a tie-breaker.
    Textual node reordering in \autoref{fig:example2-connection-good} prevents the dangling node in \autoref{fig:example2-connection-tie}.
    \autoref{fig:example2-connection-problem} uses an adjusted node order.
    }
    \label{fig:example2-connection}
\end{figure}

The developer may reorder states without changing the semantics of the model.
However, the initial state, \ie, the state \textsff{Start}, is usually defined at the top of a textual model.
Moreover, the textual SCCharts model employs secondary notation such that the node model order indicates the control-flow direction.
Additionally, states between a common start and end state are typically grouped based on the states' \emph{control-flow branch} from start to end.
\Eg, the textual model for \autoref{fig:example2-connection-good} defines the states \textsff{Ok} and \textsff{Active} that belong to the same control-flow branch below the state \textsff{Failed}, \textsff{Disconnected} at the start, the states \textsff{Inactive} and \textsff{Disconnect} at the end.
Hence, we argue that the order of connected nodes is intentional and should shape the flow of the diagram (R1).

The edges or transitions cannot be reordered without changing the semantic of a model since their textual ordering determines the transition priority.
Only mutually exclusive transition guards allow to freely reorder the transitions on a state without changing the behavior.
Hence, if the ordering of model elements has semantics, they cannot be freely reordered to exert control (R1).
The diagram depicts the priority as a number, \eg, \textsff{Start} has two outgoing edges, one with priority 1 and one with priority 2.
A user wants to read through the different options to go to a new state in the order that they might be taken as it is the case in the textual model.
Therefore, SCCharts users what to visualize transition priority if doing so does not compromise readability (R2).
Note that node model order determines the global edge model order, since edges are defined under a corresponding node.
Hence, edges originating from different nodes cannot be directly compared by model order.
Backward edges that are defined on a node with a higher node model order, however, have to be somehow compared to edges originating from their target nodes.
We solve this by initially placing backward edges below (or right, depending on the layout direction) forward edges to increase stability by using consistent layout decisions if no obviously better alternative exists, as seen in \autoref{fig:example2-connection-problem}.

Note that the SCCharts edge model order implicitly defines the port model order since ports are not explicit in the textual syntax but are only created as part of the graph representation.
Therefore, the implicitly defined port model order should not control the layout.

Previous work on model order for SCCharts~\cite{DomroesvH22,DomroesRvH23} resulted in the following model order configurations.
Strategy 1 uses the strict model order cycle breaker that enforces the node model order along edges since the node model order employs intended secondary notation.
Since the transition priorities are important to developers but could compromise the layout, Strategy 1 uses the edge model order to pre-order node and edges before the crossing minimization step, which may revise the pre-order to reduce edge crossings.
Additionally, Strategy 1 uses the node and edge model order as secondary criterion for crossing minimization such that the solution with minimal crossings that respects most of the model order will be chosen if the initial ordering is not crossing minimal.
Strategy 1 was already field-tested while teaching two students courses about Embedded System and Synchronous Languages and is currently also employed by \ac{sb}.

Strategy 2 has a different use-case and fully controls the layout by model order to create desired layouts for documentation or presentations.
The strategy uses the node model order to control cycle breaking as also done by Strategy 1 and uses the edge model order to completely control the crossing minimization step.
Strategy 2 omits crossing minimization via layer sweeps but only uses the model order to order nodes and edges as a pre-processor~\cite{DomroesvH22}.

Additionally to our own experience with SCCharts, we interviewed five \ac{sb} developers about 36 SCCharts models created by \ac{sb} for non-safety critical projects in the railway domain.
The biggest regions in all models range from 3 to 11 states with a mean of 6.22 states and a median of 6, with a mean degree of 1.65 without self-transitions.
Hence, most of the models have few nodes and edges.
The biggest model is an outlier and has in its biggest region 10 states and 43 edges without self-transitions and is regarded as large and very complex by \ac{sb} developers.
Of these 36 models, 32 have a planar embedding.
For 19 models, Strategy 1 and 2 created identical layouts.
Eleven additional models could be laid out the same way if one or two edges would be reordered.
Hence, most models intuitively employ model order, which provides control and stability.
The reasons why Strategy 2 could not directly create the same drawing as Strategy 1 falls in three patterns:
1) Backward edges that are drawn below forward edges, 2) dummy node problems, which are nodes that determine the routes of edges, have no model order, and 3) the one highly connected model is too complex to let humans manually reorder edges.

The problem of backwards edges and model order can be seen in \autoref{fig:example2-connection-problem}.
The edge from \textsff{Active} to \textsff{Interrupted} has a higher model order than the edges between \textsff{Interrupted} and \textsff{Requested} and is, hence, layouted below, as seen in \autoref{fig:example2-connection-problem}.
Since SCCharts declares their edges after their nodes, the edge model order of backward edges will always be lower than the edge model order of forward edges if the cycle breaking will be done by model order.
Hence, drawing backward edges below other incoming edges may create edge crossings if crossing minimization is omitted.

\begin{figure}
    \centering
    \includegraphics[scale=0.5]{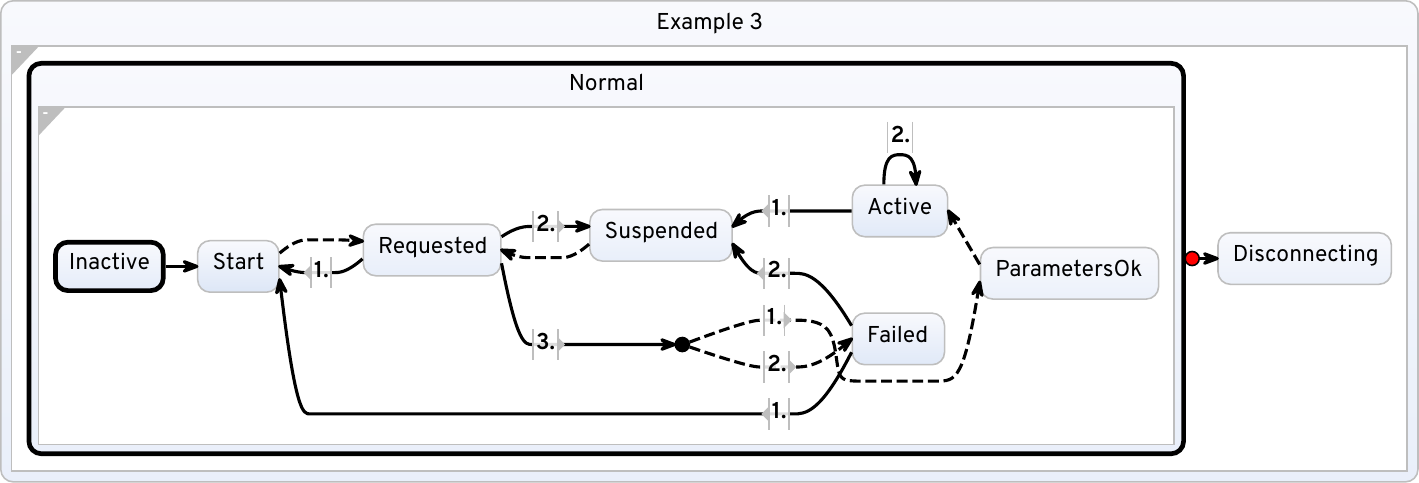}
    \caption{An obfuscated SCCharts model created by and used with permission of \acl{sb}.
    }
    \label{fig:example3-multisource}
\end{figure}

Another problem caused by strict model order crossing minimization can be seen in \autoref{fig:example3-multisource}.
Dummy nodes can only be compared to real nodes using the connection to the previous layer~\cite{DomroesRvH23}.
In \autoref{fig:example3-multisource}, an \ac{ono} layout occurs since the first connection of \textsff{Failure} to \textsff{Suspended} is used to decide whether to route the edge below or above \textsff{Failed}.
We should consider all connections of \textsff{Failed} instead or do crossing minimization to solve this issue.

On no occasion, developers or students reported anything out of the ordinary or problems as a result of the cycle breaking step controlled by model order, \eg, backward edges that should not be there.
Further analysis showed that two \ac{sb} models had unnecessary backward edges.
Developer interviews revealed that the backward edges were neither explicitly intended by the developers nor did they see them as a problem.
\ac{sb} developers did not notice that the node model order constrained the flow of the diagram before pointing it out via \autoref{fig:example2-connection-tie}.
The connector state (black dot) above the initial state \textsff{Disconnected} is a dangling node---a node with only edges to the right---even though it is not a source node.
\autoref{fig:example2-connection-good} solves this issue but expresses a different secondary notation.
The layout does no longer suggest that the two ways to the \textsff{Requested} state are the \textsff{Interrupted} and \textsff{Renewed} states and hides their symmetry.

\ac{sb} developers use the diagram mainly to browse the model and for documentation as well as communication with domain experts.
Typically, they edit only small details and only open the diagram to verify their changes.
This may affect how they utilize model order compared to students who open text and diagram side-by-side while editing.
When comparing the models by \ac{sb} developers to student models, this does not seem to affect the model order when comparing models of similar size.
\ac{sb} models, however, did have no occasions of \emph{copy-paste anomalies} where the model order was completely unintentional by the fact that states and transitions were copied without considering their intention.

Generally, \ac{sb} developers preferred diagrams with fewer edge crossings over drawings that enforced model order during crossing minimization as employed by Strategy 2.
Nevertheless, they still want to use model order as a tie-breaker to increase stability (R2), which Strategy 1 provides for them.

The developers reported that transition guards are usually mutually exclusive, which would allow to reorder the transitions to control the layout (R1).
Browsing through the different models, however, showed exceptions including cases that could not be verified since the guards employed host code functions.
Conservatively, we should assume that the edge order cannot be freely controlled (R1).

To summarize, the strict model order cycle breaking is active as the default layout for about a year, which allows us to gather feedback on the used layout options, and did not produce \ac{ono} layouts or left developers confused.
Hence, it should remain the default option to control the flow via model order (R1).
Note that this might not be the case if a language uses a textual syntax that only models edges or edges consisting of source and target instead of edge being defined below their source node.

Strategy 2 cannot be the default option for SCCharts, since it produces \ac{ono} layouts.
However, as a second strategy, we still need a strategy that can exert more control to create exactly the topology we are envisioning (R1).
Since the full control strategy produces \ac{ono} layouts, we should investigate whether a greedy post-processing of strict model order crossing minimization or only constraining the order of SCCharts states can provide enough control to influence the topology without creating \ac{ono} layouts.

A controlling, more strict, strategy has additional uses.
If the textual model controls most parts of the layout, a bad layout often points to a bad textual model.
\Eg, if the initial state is somewhere in the middle of the diagram, it might be somewhere in the middle and not at the start of the textual model.
We, however, currently do not have enough data or feedback for the full control variants to confirm or disconfirm this hypothesis.

\subsection{Controlling Lingua Franca Layout via Model Order}

\begin{figure}
	\centering
    \begin{subfigure}[b]{0.32\textwidth}
        \centering
        \begin{lstlisting}[language=LF, escapeinside={(*}{*)}, abovecaptionskip=0pt, captionpos=t, basicstyle=\sffamily\scriptsize,]
main reactor {
    a = new Accelerometer();
    dx = new Display(row = 0);
    dy = new Display(row = 1);
    timer t(0, 250 msec);
    state count:int(0);
    reaction(t) -> a.trigger (*$\dots$*)
}
        \end{lstlisting}
        \vspace{-2em}
        \caption{}
        \label{fig:lf-example-text}
    \end{subfigure}
    \begin{subfigure}[b]{0.65\textwidth}
        \centering
        \includegraphics[scale=0.5]{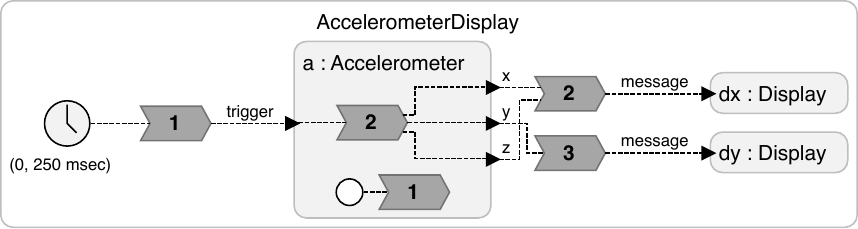}
        \caption{}
        \label{fig:lf-example-graph}
    \end{subfigure}
	\caption{
        The \textsff{AccelerometerDisplay} \ac{lf} example model with abbreviated textual model and graphical representation.
    }
	\label{fig:lf-example}
\end{figure}

\acl{lf} \cite{LohstrohMBL21} is a polyglot coordination language for reactive real-time systems that models data-flow.
\ac{lf} developers typically use the diagram together with the textual model to utilize the advantages of both representations, since the textual model alone can be tedious to work with.
An example \ac{lf} model can be seen in \autoref{fig:lf-example}.

\autoref{fig:lf-example-text} depicts the main reactor of the \ac{lf} model and \autoref{fig:lf-example-graph} depicts the corresponding diagram with a timer (clock), reactions (numbered arrows shapes), and reactors (gray boxes).
Users may define states, actions, timers, reactors, edges between reactors, and reactions and often define the components in the mentioned order forming \emph{ordering groups}.
\Eg, users define reactors in a separate group above the ordering group of reactions.
Since the language does not constrain the order, some models diverge from this schema, as seen in \autoref{fig:lf-example-text} which defines reactors first.
It would even be possible interleave reactors and reactions in the textual model.
A reactor may again consist of everything the main reactor has to offer but may define inputs and outputs, which are typically at the top of the reactor in the textual model to express intention.
Edges between reactors and other elements are created implicitly based on their interfaces, \ie, their declared inputs and outputs, which corresponds to the port model order.
\Eg, the reaction uses the timer \textsff{t} as input and outputs \textsff{a.trigger}, a trigger of the \textsff{Accelerometer} \textsff{a}.

Reactors have two model orders they could adhere to: Their definition order, which corresponds to the place \textsff{Accelerometer} and \textsff{Display} are defined, or their instantiation order, which corresponds to the creation of the reactor instances.
We use the instantiation order as the node model order, since the definition order of multiple instances of a reactor is the same and another file may hold the reactor definition.
The reactor instantiation order can be freely changed and typically expresses desired secondary notation that could be controlled via model order (R1).
\Eg, drawing the \textsff{Accelerometer} before the two \textsff{Display} reactors visualizes the data-flow and the order in the textual model.

The order of reactions defines their scheduling order.
Hence, we cannot freely reorder them to control the layout (R1).
However, this scheduling order is something that developers want to represent in their drawing, similar to SCCharts edge priorities.
At the first introduction of diagrams for \ac{lf}, reactions could be freely placed without considering their order.
\ac{lf} developers, however, found it irritating that the reaction numbers were not in order.
Hence, ordering the reactions in the diagram by model order is not a constraint but a deliberate choice to visualize part of the reaction semantics.
Conversely, the language does not constrain the order of reactors, reactor edges, actions, timers, states, inputs or outputs, they only adhere to ordering conventions between the ordering groups.

The ordering of the reactors, reactions, actions, and timers is only relevant in their respective ordering group.
Model order can only directly compare reactors to reactors but not explicitly defined edges between reactors to implicitly created edges between actions and reactions.

Interviews and presentations as part of the weekly \ac{lf} team meeting as well as interviews with Magnition that want to use \ac{lf} to model cache structures resulted in the following statements by developers.
\begin{enumerate}
    \item Edge-crossings should be reduced.
    \item Stability is important for large hierarchical models.
    \item Drawings are generally too wide.
    \item The ports of reactors define their interface and should hence be respected.
    \item The reactor-instantiation-order could be used to control their ordering.
    \item Reactions and actions should be drawn below reactors since reactors are more important.
    \item Actions should be placed such that the flow indicates how they may be triggered.
\end{enumerate}
Note that statement 1, 2, and 3 came from Magnition developers and 1, 3, 4, 5, 6 from the person in the \ac{lf} community that mainly interacted with us during the presentations and discussions.
All the statements above were additionally verified by one-on-one interviews with a small group of developers by showing and asking question regarding the layout of \ac{lf} models.
The key insights were gained by not showing perfect layouts but by showing layouts that could be improved to give the developers incentive to respond.

Statement 1 is an established aesthetic criterion that reduces clutter and makes drawings more readable (R2).
Reducing edge crossings conflicts with constraining the order of nodes or ports.
Hence, edge crossings and stability are conflicting, as seen in \autoref{fig:cdn-cache-snippet}.

\ac{lf} uses cross-hierarchy edges, as seen in \autoref{fig:lf-example}.
The edge between reaction 2 inside the \textsff{Accelerometer} to the reactions 2 and 3 inside the \textsff{AccelerometerDisplay} goes through the node of the \textsff{Accelerometer} reactor.
Since we layout the inner graph inside the \textsff{Accelerometer} with its three ports to the upper hierarchy level first, the ports \textsff{x}, \textsff{y}, and \textsff{z} have a predetermined order as a result of the output order of the reaction 2 inside \textsff{Accelerometer}.
When layouting the \textsff{AccelerometerDisplay}, the port positions of \textsff{Accelerometer} must, therefore, remain stable as requested in statement 2 and 4\footnote{Note that layout strategies could minimize the edge crossings as a result of different hierarchy level, \eg by changing the order of the ports to \textsff{x}, \textsff{z}, \textsff{y}, but these do not guarantee stability}.
In highly connected models such as the \textsff{Pop} reactor depicted in \autoref{fig:cdn-cache-snippet}, changing the port order may drastically change the layout.
If reactors are again used in other reactors as it is the case for \textsff{Pop}, small changes in a model may lead to large changes in the diagram, which compromise the mental map.

Concerning Statement 3, the horizontal labels and the layout direction from left-to-right results in drawings that are generally wider than high and results in unfavorable aspect ratios to display the diagram side-by-side with the textual model.
Hence, developers typically place textual model and diagram under each other instead of side-by-side.
The width of a drawing partly also depends on the cycle breaking algorithm in use.
More backward edges often result in fewer layers and a higher drawing.
Since \enquote{having fewer layers} stands in conflict with the minimal backward edges aesthetic criterion used for cycle breaking, the user should express this explicitly if desired, \eg, by the reactor instantiation order (statement 5).

Statement 6, drawing reactions and actions below reactors, may again conflict with minimizing the edge crossings and preserving the reactor port order.
Additional interviews revealed that this does not depend on the nature of the semantic elements but rather whether small elements (reactions, actions, startup) might be split up by placing a big reactor between them, which is undesired.

Statement 7 reveals that not only the structural flow but also the nature of the different semantic elements may be important.
Hence, a layout strategy should be able to control the placement of semantic elements relative to other semantic elements (R1) to be able to visualize how to enable actions.

We conclude that we should not use model order strategies that completely constrain the layout for \ac{lf} (R1).
Layout algorithms for \ac{lf} should use model order only as a tie-breaker as the default strategy for cycle breaking and crossing minimization, and may constrain nodes of the same ordering group to avoid misleading secondary notation.
For cycle breaking the depth-first strategy proved to be good to handle the often intertwined action and reaction networks quite well.
The depth-first algorithm itself provides more stability than the currently employed greedy cycle breaker that optimizes backward edges.
The depth-first model order cycle breaker uses the node model order (or any other model order) as a tie-breaker to determine the visiting order of nodes during depth-first search.
However, depth-first cycle breaking has the disadvantage that it creates wide layouts compared to other approaches.
If graphs remain relatively small as it is the case for \ac{lf}, a cycle breaking algorithm based on strongly connected components that utilizes model order to reverse edges in each component separately may create better and more compact drawings.
As an additional strategy, strict model order cycle breaking~\cite{DomroesRvH23} gives the developer full control over the flow of the diagram.
Since a user can reorder reactors, reactions (at least partly), actions, and timers by disregarding the ordering conventions, this strategy can create desired drawings for documentation or presentations (R1).

There is no obviously optimal crossing minimization strategy for \ac{lf}.
As a default strategy, model order should only be considered between elements of the same ordering group, \eg, comparing reactors to reactors by model order as a new criterion for the barycenter heuristic for crossing minimization~\cite{DomroesRvH23}.
Crossing minimization may also constrain the port order and the node order for big hierarchical models, since this creates stability and preserves the mental map by using the interface of reactors.
Additionally, \ac{lf} requires a model order crossing minimization strategy to control the layout if necessary, which requires to break the ordering conventions to create desired layouts.

\section{Gathering Feedback from Developers}
\label{sec:feedback-gathering}

In our experiences over many years, the first and hardest step of gathering feedback from developers is finding enough developers willing to cooperate that employ the language for real use-cases.
The second problem is finding enough models with a real use-case to analyze since these are often classified.

SCCharts and \ac{lf} are used in research, teaching, and industry.
Regular students typically have beginner knowledge of the language and create unrealistic models of relatively small size as part of weekly assignments.
Researchers create models to showcase specific language features or a specific problem with the language or solved by the language, as well as test-cases for the language.
Software developers that use the language for a real-world application create realistic models of realistic size and complexity.

Quantitative analyses of the models does typically not work since the number of models and the number of developers is very small.
For Lingua Franca, we had access to the \textsff{cdn\_cache\_demo} model already mentioned above, and to several example \ac{lf} models created by researchers, which are a lot smaller and a lot less complex, leaving us with only one realistic model.
In our experience, recording the editing process---as employed by Purchase \etal~\cite{PurchasePP12}---for real-world models is difficult since models may be classified and are typically created over several weeks or month.
What remains are qualitative interviews that have the disadvantage that demonstrating new tooling might influence the participants and leading questions can occur.
Additionally, getting feedback from developers requires follow-up questions to understand what they want and interactive tooling to show different alternatives.
Hence, a quantitative analysis, which would have involved ranking layouts of the same graphs, does not seem promising since the process of layout creation and the intention behind textual ordering is lost.

To gather feedback for SCCharts and \ac{lf}, we first analyzed existing models, which requires to know the language and the editing workflow in question to identify anomalies and use-cases.
As a second step, we searched and found patterns that indicated intention in the textual model that we could use to control the layout to improve the secondary notation.
Based on that, we presented to and interviewed developers showing possible options to exert control via model order to extract which ordering in the textual model is intended secondary notation and how it should be presented in the diagram.

Doing these interviews revealed additional obstacles.
In our experience developers that do not know automatic layout and how it works have difficulties giving feedback for layout questions and tend to mix it with visualization aspects.
Developers without prior knowledge cannot balance between edge crossings and stability but often want both, which is not always possible.
Moreover, untrained developers do tend to mix up the ordering in layout direction, which determines the direction of edges and is a result of the cycle breaking step, with the ordering orthogonal to the layout direction, which is determined by the crossing minimization step.
This gets aggravated if one considers that a different set of reversed edges for cycle breaking might result in more or fewer nodes in a layer, which does hence indirectly influence the outcome of crossing minimization.
Hence, feedback for cycle breaking and crossing minimization should not be gathered at the same time.

On several occasions we got feedback that developers wanted more options to configure the layout.
However, this does only work if they already know the underlying layout algorithms.
We noted that too many layout configurations tend to confuse developers.
They will only use options to configure the diagram that they know and understand.

Different developer teams might employ different secondary notation.
We asked developers from Magnition and in the \ac{lf} weekly meetings regarding ordering and got partially different answers.
This might be the case because they employ the language for different use-cases and employ different secondary-notation based on previously used tooling.
This might also be the case for different SCCharts developing teams.

\section{Conclusion}
\label{sec:conclusion}

How SCCharts and \ac{lf} layout can be controlled by model order can be generalized for similar languages.
Most insights about SCCharts can be generalized for state-machines that follow a similar state and transition declaration pattern.
Hence, we suggest to constrain the flow of all state machines by the state model order to create intentional backward edges.
Data-flow languages similar \ac{lf} might want to use a mix of semantics and model order to determine which edges to reverse since the ordering may be constraint by the flow of information.

Model order should only be used to control the layout and used as a constraint if the underlying textual ordering matches the user intention and can be controlled by the user.
\Eg, the ports in SCCharts diagrams are not explicitly modeled but implicitly created by edge order.
Hence, the edge order should influence the order of edges and not the ports order hence it cannot be directly manipulated.
For \ac{lf} or other languages that explicitly models ports, this is the other way around.

Model elements that are ordered by convention or restrictions in the grammar would only control the diagram if these conventions or restrictions express intention.
If this is not the case, the ordering should not be used to constrain the layout but can only be used as a tie-breaker.
\Eg, the order of SCCharts edges corresponds to their priority, which should be visualized.
A similar constraint exist for the reaction execution order in \ac{lf}.
The order of a reactor instantiation and a reaction is determined by convention not by intention and should hence not constrain their positioning.
This fine-tuning of layout for a given language does, hence, require the ability to differentiate semantic elements in the layout step and a layout algorithm that is able to create individual constraints or layout preferences for model elements or groups of model elements.
Future work on this should investigate whether numbering of model elements creates the desire to see them ordered, which could be further leveraged to extract intention from a given model that goes beyond textual ordering.

While flow of information can often be constrained to create secondary notation that visualizes how data or control may go backwards, constraining crossing minimization by model order should not be a default strategy for automatic layout.
In many cases, this may create \ac{ono} layouts and may increase the edge crossings significantly.
Nevertheless, the control given by strict model order strategies is desired to create specific layouts for documentation or presentations.
\Eg, SCCharts priorities are an important part of the model, however, constraining edges based on the edge model order constrains leaves no leverage to optimize edge crossings and is hence undesired as a default option.
Good layouts created by strict model order crossing minimization can also be created using model order as a tie-breaker.

If the underlying graph of a visualization is planar, the textual order of human-made real-world control-flow models very often conforms with the model order.
This is the case since state-machines form branches of control-flow from start to end or repeat.
If the model does not have any connections between the branches, the layout becomes trivial and conforms to the model order.

Furthermore, for languages with cross-hierarchical edges and expandable and collapsible hierarchical nodes, stability is a very important aspect, sometimes more than edge crossings if the model is often edited or elements change their port order after being expanded or collapsed.
A deterministic port order controlled by the port model order can solve this problem.
This should be further investigated with additional models and realistic tasks in an interactive setting.

%
%
%
\bibliographystyle{splncs04}
\bibliography{local.bib}
\end{document}